\documentclass[a4paper]{jpconf}
\usepackage{amssymb,graphicx}

\begin{document}

\title{Collective excitations in neutron-star crusts}

\author{N Chamel$^1$, D Page$^2$ and S Reddy$^3$}

\address{$^1$ Institut d'Astronomie et d'Astrophysique, Universit\'e Libre de Bruxelles - CP226, 1050 Brussels,  Belgium}
\address{$^2$ Instituto de Astronom\'ia, Universidad Nacional Auton\'oma de M\'exico,  Mexico D.F. 04510, Mexico}
\address{$^3$ Institute for Nuclear Theory, University of Washington, Seattle, Washington 98195}

\ead{nchamel@ulb.ac.be}

\begin{abstract}
We explore the spectrum of low-energy collective excitations in the crust of a neutron star, especially
in the inner region where neutron-proton clusters are immersed in a sea of superfluid neutrons. The speeds of the 
different modes are calculated systematically from the nuclear energy density functional theory using 
a Skyrme functional fitted to essentially all experimental atomic mass data. 
\end{abstract}

\section{Introduction}

A few meters below the surface of a neutron star at densities above 
$\sim 10^4$ g$\cdot$cm$^{-3}$, atomic nuclei are fully ionized by the pressure thus coexisting with a highly degenerate 
electron gas. As the density increases, nuclei become more and more neutron-rich due to electron captures~\cite{pearson2011} 
until neutrons start to drip out of nuclei at densities above $\sim 4.10^{11}$ g$\cdot$cm$^{-3}$~\cite{pearson2012} thus 
delimiting the boundary between the outer and inner parts of the crust. Those ``free'' neutrons are expected to become 
superfluid below a critical temperature of the order of $\sim 10^{10}$~K. Despite the absence of viscous drag, the neutron 
superfluid can still be strongly coupled to the crust due to non-dissipative entrainment effects~\cite{chamel2012}.

Although the crust of a neutron star represents only about $1\%$ of the star's mass, it is expected to play a key role 
in various observed astrophysical phenomena~\cite{lrr}. In particular, the collective excitations in the inner crust of a neutron star 
can impact the thermal evolution of the star~\cite{pr12}. These excitations have been already studied in the Wigner-Seitz approximation, using 
self-consistent mean-field methods with Skyrme effective interactions~\cite{grasso2008,baroni2010}. In this approach, 
the crust is divided into a set of independent spheres centered around each nucleus. Consequently, only high-energy 
collective excitations with wavelengths smaller than the size of the Wigner-Seitz cell can be studied in this framework. 
Moreover, this approximate treatment of the crust does not take into account entrainment effects, which arise from Bragg 
scattering of unbound neutrons by the Coulomb lattice. For this reason, we have recently followed a different approach 
by employing the band theory of solids. In this way, we have been able to study low-energy collective modes including 
entrainment effects~\cite{chamel2013}. 

In this paper, we extend our previous study of collective modes by including the coupling of the neutron superfluid to the strain 
field. 

\section{Elasto-hydrodynamics of the neutron-star inner crust}

The normal modes of oscillations of the neutron superfluid permeating the neutron-star inner crust can be obtained by linearizing 
the elasto-hydrodynamic equations (see, e.g., Refs.~\cite{pet10,kob13}). Treating the crust as an isotropic solid, the normal modes consists of 
two transverse modes and two longitudinal modes. Using the same notations as in Ref.~\cite{chamel2013}, the speed of the transverse 
modes is given by $v_t =\sqrt{S/\rho_{\rm I}}$, where $S$ is the shear modulus of the crust and $\rho_{\rm I}$ is the mass density associated with lattice 
vibrations. Because unbound neutrons are entrained by the crust, the mass density is much larger than the ion mass density, and is given by 
$\rho_{\rm I}=m (n_p+n_n^{\rm b})$, where $m$ is the nucleon mass, $n_p$ the proton density and $n_n^{\rm b}$ the density of entrained neutrons. 
In other words, including entrainment significantly reduces the values of $v_t$. The longitudinal modes satisfy a dispersion relation for two 
coupled modes, namely 
\begin{equation}\label{1}
(v^2-v_\phi^2)(v^2-v_\ell^2)=g^2_{\rm mix}v^2 +g^4\, ,
\end{equation}
where $v_\phi$ is the speed of the Bogoliubov-Anderson bosons of the neutron superfluid, and $v_\ell$ is the speed of longitudinal lattice phonons. 
The mixing between these two modes is characterized by the parameters 
$g_{\rm mix}$ and $g$, given by 
\begin{equation}\label{2}
g_{\rm mix}=\sqrt{\frac{n_n^{\rm b} (2 L+E_{nn} n_n^{\rm b})}{\rho_{\rm I}}}\, , \hskip 0.5cm g=\left(\frac{L^2 n_n^{\rm c}}{m\rho_{\rm I}}\right)^{1/4}\, ,
\end{equation}
where $L=n_p \partial \mu_n/\partial n_p$ ($\mu_n$ being the neutron chemical potential and $n_n$ 
the neutron density), $E_{nn}=\partial\mu_n/\partial n_n$, and $n_n^{\rm c}=n_n-n_n^{\rm b}$ is the density of ``conduction'' neutrons. Note that 
Eq.~(\ref{1}) is identical to Eq.~(34) from Ref.~\cite{kob13} although it is expressed here in a slightly different form. 

In our previous study~\cite{chamel2013}, we assumed that the coupling of the superfluid to the strain field embedded in the coefficient $L$ is small, 
and we thus set $L=0$. We now include this coefficient. The two solutions of Eq.~(\ref{1}) are given by 
\begin{equation}\label{3}
 v_{\pm} = \frac{V}{\sqrt{2}}\sqrt{1\pm\sqrt{1-\frac{4 v_\ell^2 v_\phi^2}{V^4}+\frac{4g^4}{V^4}}}\, ,
\end{equation}
where $V = \sqrt{v_\ell^2 + v_\phi^2 +g_{\rm mix}^2}$. 

In the non-superfluid phase, any relative motion between the neutron liquid and the crust will be damped by viscosity to the effect that 
ions, electrons and neutrons will be essentially comoving. Only one longitudinal mode corresponding to ordinary 
hydrodynamic sound will persist and its speed will be given by $c_s =\sqrt{(K+4S/3)/\rho}\approx\sqrt{K/\rho}$ where $\rho$ is the total mass density of the crust 
and $K$ is the total bulk modulus, which is related to the bulk modulus $\widetilde K$ of the electron-ion system by 
$K=\widetilde K + 2 n_n L + n_n^2 E_{nn}$. Assuming that pairing affects neither the composition nor the elastic properties of the crust, 
the coefficient $L$ can be obtained by inverting the previous equation, thus yielding 
\begin{equation}\label{4}
 2 n_n L=K- \widetilde K-n_n^2 E_{nn}\approx \rho c_s^2-\rho_{\rm I} v_\ell^2- \frac{m n_n^2 v_\phi^2}{n_n^{\rm c}}\, .
\end{equation}

\section{Microscopic model for the inner crust of a neutron star}  

As in our previous work~\cite{chamel2013}, we use the inner crust composition given in Ref.~\cite{onsi08}, which was obtained from the 
fourth-order extended Thomas-Fermi method with proton quantum shell effects added via the Strutinsky-Integral theorem. This so-called ETFSI method 
is a high-speed approximation to the self-consistent Hartree-Fock equations. The calculations of Ref.~\cite{onsi08} were carried out using 
the BSk14 Skyrme interaction underlying the HFB-14 atomic mass model~\cite{sg07}, which yields an excellent fit to essentially 
all experimental atomic mass data with a root mean square deviation of 0.73 MeV. Moreover, the BSk14 interaction was also constrained to 
reproduce a realistic neutron-matter equation of state. 
For these reasons, this interaction is particularly well-suited for describing the inner crust of a neutron star. 

\begin{figure}[h]
\begin{center}
\includegraphics[width=10cm]{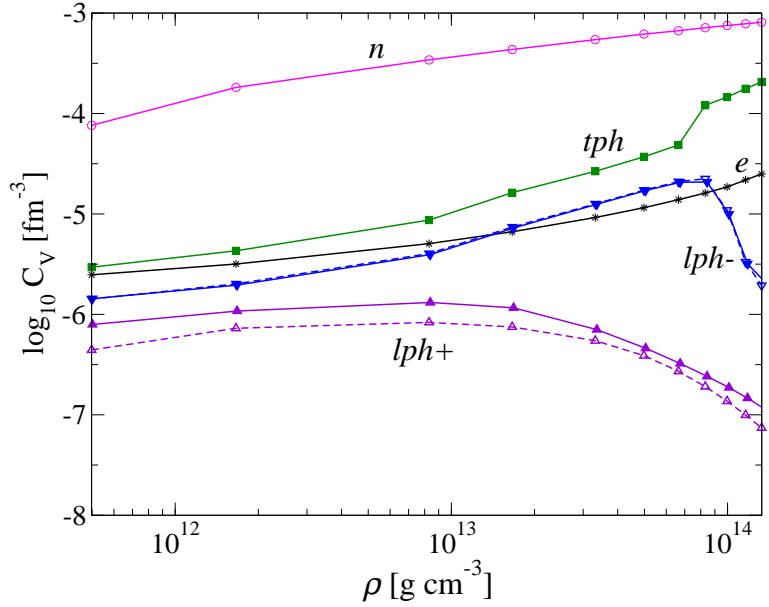}
\end{center}
\caption{Heat capacity of electrons ($e$), transverse lattice phonons ($tph$) and longitudinal excitations ($lph-$ and $lph+$) 
in the inner crust of neutron stars at temperature $T=10^9$~K, with the coupling of the superfluid to the strain field (solid lines) and without (dashed lines). 
For comparison, is also shown the normal neutron contribution ($n$), but it is strongly suppressed by superfluidity except
in the shallowest and densest parts of the inner crust where the neutron $^1$S$_0$ pairing gap becomes vanishingly small.}
\label{fig1}
\end{figure}

Although the neutron quantum shell effects are much smaller than proton ones~\cite{oya94}, and therefore are expected to have a small impact on the crustal 
composition, they drastically affect the entrainment between the neutron superfluid and the crust. Neutron quantum shell effects were systematically studied 
in Ref.~\cite{chamel2012} in the framework of the band theory of solids using the mean-field potentials obtained self-consistently from the ETFSI method in 
Ref.~\cite{onsi08}. This allowed the calculation of the conduction neutron density $n_n^{\rm c}$. The quantities $E_{nn}$, $v_\phi$, $v_\ell$ 
and $c_s$ are calculated as in Ref.~\cite{chamel2013}. The $L$ coefficient is then estimated from Eq.~(\ref{4}). The numerical results are shown in Table~\ref{tab}. 
It can be seen that the coupling of the superfluid to the strain field reduces the speed of the highest longitudinal mixed mode, whereas the speed of the lowest 
longitudinal mixed mode is barely affected. Since the contribution 
of a collective excitation with velocity $v$ to the heat capacity varies like $v^{-3}$ at low temperatures, taking the coefficient $L$ into account 
substantially increases the heat capacity of the highest longitudinal mixed mode, which is enhanced by about a factor of $3$ at temperature $T=10^7$~K near the neutron-drip 
transition, as compared to our previous estimates. However, the contribution of this mode to the crustal heat capacity still remains negligible, as shown 
in Fig.~\ref{fig1}. Likewise, the thermal conductivity of the crust remains almost unchanged compared to our previous estimates. Indeed, 
the electron-phonon scattering rate, which scales like $v^{-3}$, is dominated by processes involving transverse phonons. 

\begin{table}[h]
\caption{\label{tab}Properties of collective modes in the inner crust of a neutron star for different values of the average baryon number density $\bar n$. 
The speeds are expressed in units of $10^{-2}c$, $c$ being the speed 
of light. Values in parenthesis were obtained by neglecting the coupling of the superfluid to the strain field. }
\begin{center}
\begin{tabular}{lllllll}
\br
$\bar n$ (fm$^{-3}$) & $v_\phi$ & $v_\ell$ & $c_s$ & $v_t$ & $v_{-}$ & $v_{+}$ \\
\mr
0.0003 & 1.11 & 5.13 & 5.35 & 0.578   & 0.935 (1.02)  & 3.75 (5.56) \\
0.001 & 1.34 & 3.69 & 4.46  & 0.416   & 1.27  (1.08)  & 3.50 (4.58) \\
0.005 & 1.64 & 2.60 & 4.78  & 0.293   & 1.01  (0.890) & 3.86 (4.80) \\
0.01 & 1.77 & 2.39 & 5.18   & 0.251   & 0.897 (0.814) & 4.40 (5.20) \\
0.02 & 1.42 & 2.26 & 5.84   & 0.237   & 0.582 (0.548) & 5.35 (5.84) \\
0.03 & 1.62 & 2.31 & 6.55   & 0.242   & 0.595 (0.570) & 6.20 (6.56) \\
0.04 & 2.18 & 2.44 & 7.39   & 0.256   & 0.750 (0.720) & 6.96 (7.39) \\
0.05 & 4.21 & 2.83 & 8.30   & 0.235   & 1.50 (1.44)   & 7.68 (8.31) \\
0.06 & 5.86 & 3.31 & 9.28   & 0.275   & 2.16 (2.09)   & 8.37 (9.29) \\
0.07 & 7.76 & 4.26 & 10.3   & 0.354   & 3.25 (3.20)   & 9.07 (10.3) \\
0.08 & 8.98 & 4.87 & 11.4   & 0.404   & 3.70 (3.84)   & 9.82 (11.4) \\
\br
\end{tabular}
\end{center}
\end{table}

\section{Conclusions}

Due to nondissipative entrainment effects, the Bogoliubov-Anderson bosons of the neutron superfluid in the inner crust of a neutron star
are strongly mixed with the longitudinal crystal lattice phonons. The coupling of the superfluid to the strain field, which we neglected in our 
previous study~\cite{chamel2013}, is found to substantially reduce the speed of the highest longitudinal mixed mode, especially in the 
shallowest layers of the inner crust. However, the thermal properties of the crust are almost unchanged.

\ack
The financial support of F.R.S.-FNRS (Belgium) is gratefully acknowledged. 
The work of S.R. was supported by DOE Grant No. \#DE-FG02-00ER41132 and by the 
Topical Collaboration to study {\it neutrinos and nucleosynthesis in hot dense matter}. 
D.P's work was supported by grants from Conacyt (Grant No. CB-2009/132400) and UNAM-DGAPA (Grant No. PAPIIT IN113211).


\end{document}